\documentclass[]{UCD_CS_FYP_Report}
\usepackage{graphicx}
\usepackage{float}
\usepackage{hyperref}

\def\studentname{Ois\'in Kyne} 
\def\projecttitle{MAC Address De-Anonymisation} 
\def\supervisorname{Dr. Brian MacNamee}

\begin{document}

\maketitle


\begin{abstract}

This project is an exploration into analysing WiFi probe requests, a management frame described as part of the IEEE 802.11 protocol which publicly broadcasts the senders MAC address. The intention was to collect these probe requests to use as a basis to link people's information to the MAC of their device. This would enable people to be identified in future by the presence of their device near a probe request detector. 

Due to data protection and privacy issues preventing access to real data of people's names and locations, this project was divided into two parts. Firstly, an identification algorithm was developed and tested on simulated data sets of MAC addresses and names, to prove MAC address identification is possible. And secondly, a distributed system of probe request detectors coupled with a centralised MAC address database was developed to demonstrate that these simulated MAC addresses are obtainable in the real world. 

The capturing software was initially developed for Unix systems and uses a Django-powered web server to store data from multiple capturing devices. Python was used to model and test the identification algorithm. 

\end{abstract}
\newpage


\chapter*{Acknowledgements}

I would like to express my gratitude to my Project Supervisor Dr. Brian MacNamee for his support and advice, as well as for providing me with both resources from his previous demonstrations on the topics of WiFi probe requests and spare hardware with which I was able to set up a permanent WiFi sniffing device.  \\

I would also like to thank Mr. Paul Dolan for his advice on wireless networks and on academic writing in general. \\


\tableofcontents\pdfbookmark[0]{Table of Contents}{toc}\newpage
\newpage


\chapter{Introduction}
Every WiFi enabled device manufactured has a unique identifier. This identifier is called a MAC address and is normally stored in read-only memory on the device to prevent it being changed. MAC addresses are the foundation for local area networks, by providing a unique name that every device in the area can be referred to by. When the WiFi protocol was developed, it was developed in such a way that these MAC addresses are broadcasted in plain text by any device that is searching for a nearby network to connect to. Many smartphone owners do not deactivate WiFi on their phones when going out in public, as a result the majority of phones in a public area are consistently broadcasting their MAC addresses every few seconds while searching for available WiFi networks to connect to\cite{freudiger}. 

Although MAC addresses are unique, they do not identify people alone, but need to be related to a secondary data set containing personal information. The process for creating these relations involved analysing both probe requests and the secondary data set for co occurrences between people and MAC addresses that would indicate a relationship between a pair.

This second aspect of this project involved developing software to detect these broadcasted MAC addresses and store them. These broadcasted requests include MAC addresses and occasionally, SSIDs (network names) of networks known to the device, these were stored along with the time and GPS location of where the detection took place. After detection it became a challenge of mining this data to profile the behaviour of the owners of these devices to generate actionable information from these detections.

Hypothetically, from analysis of probe requests alone, the information we could learn about a given user can include; information about their habits at a place where we are detecting probe requests (E.g. When they usually arrive/leave.), we can make an educated guess about the device we are detecting, if their device is broadcasting 'directed' probe requests it means their device is broadcasting a list of SSIDs of previously used networks, by using open source data from Wigle.net\cite{wigle} we can map the location of these networks and the likelihood that one of these SSIDs are the home networks of these users is very high. So now we know the target user's habits, the locations they frequent often enough to connect to the WiFi there (which most likely includes their home address), the manufacturer of their device and the times they show up to our known detection location. 

This is a significant amount of data and can be used to profile many people but it does need to be mixed with a secondary data set to try and put a name or contact to the MAC address to maximise the impact of the data. The user identification aspect of this project examined the conditions required and the accuracy obtained when trying to identify people based on a correlation between a probe request data set and another (in this case randomly generated) dataset that included personal information. Both the collection and analysis of probe requests and the identification process will be covered in further detail throughout the remainder of this report.


\chapter{Background Research}
In the planning and early stages of this project a large amount of research was been made into the relevant technologies and approaches required. This section will look at capturing WiFi data and the processes involved in identifying people with this data.

\section{Capturing WiFi Data}
The 802.11 specifications\cite{802.11} standardise a wireless local area connection format. First released in 1997 there have been a number of significant amendments since, mainly in the area of improving bandwidth and range. The particular parts of the standard under scrutiny in this project are the probe request packets, the probe response packets and the beacon frames that are used in the detection of nearby wireless networks. 

\subsection{802.11 Management Frames}

The part of the 802.11 specification this project looks into is the management frames responsible for beginning the process of connecting to a Wireless network. The three frames in particular we look at are the Probe Requests, Probe Responses, and Beacon Frames. When a WiFi enabled device is not connected to an access point and wants to connect to one, it sends out a probe request on each 	WiFi channel and gathers the responses from the various available access points from which to choose. These are called probe responses. 

\begin{figure}[h]
	\includegraphics[width=\linewidth]{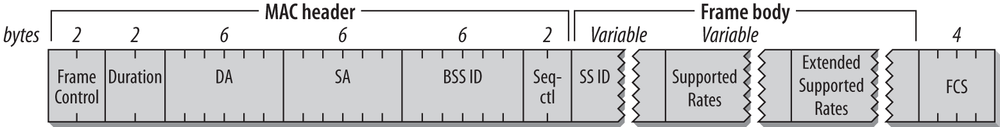}
	\caption{The layout of a probe request frame.\cite{gast}}
	\label{fig:probe}
\end{figure}

If a device wants to search for only a specific access point it can send a \emph{directed} probe request. In a directed probe request the device puts the SSID of the required access point into the probe request's SSID field (see Fig. \ref{fig:probe}); in an undirected probe request the SSID field is left blank. \\
\\
When an access point receives a probe request, it responds to the request detailing its configuration. 	It includes important details such as the encryption method the access point uses, the rates it can 	transfer data  at and its SSID; this packet is known as a probe response. This information is also 	periodically sent out in the form of a beacon frame such that devices can connect to a network passively, beacon frames are broadcasted approximately every 100ms, which requires a device's wireless radio to be active for much longer than a probe request/response cycle (approx 5ms), most portable devices rely on probes to conserve power and improve battery life. 

\subsection{Wireless Sniffing}
Wireless sniffing is the use of software to detect and log packets being passed across a network. It is commonly used to debug problematic issues with networks and to analyse network usage. Other uses include room occupancy estimates, spying, and hacking unsuspecting users on unencrypted connections.

\subsection{\label{monitor}Promiscuous Mode}
All packets using the 802.11 headers have a sender and a receiver MAC address. A network card 	will only relay a packet to the operating system if the receiver's address is set as either the broadcast address (ff:ff:ff:ff:ff:ff) or the network card's MAC address. Meaning only broadcast messages and 	messages directed to that device get passed to the OS. Promiscuous mode (sometimes called monitor 	mode) overrides this filtering and sends all packets the network card receives to the operating 	system. This allows our sniffing software to detect probe responses being sent to other devices after their probe requests.

\subsection{WiFi Channels}
The channels WiFi can be used across have been modified as the technology has progressed; at this 	point the main IEEE standards in use are called 802.11 a/b/g/n. The two main 	bandwidths that WiFi networks operate on are located at 2.4GHz and 5GHz on the electromagnetic spectrum. 

\begin{figure}[h]
	\includegraphics[width=\linewidth]{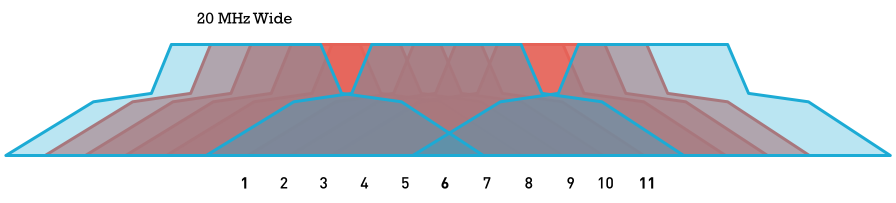}
	\caption{WiFi Channels and their overlap\cite{metageek}}
	\label{fig:WifiChannels}
\end{figure}

On the 2.4GHz spectrum there are 11 channels. Each channel's centre is spaced 5MHz apart but each 	channel has an acceptable spread of 20MHz. This means that channels overlap and interfere with 	one another. To avoid these overlapping interferences, the general consensus is to only use channels 	1, 6 and 11. As with 20 MHz spreads, these still don't overlap. Figure \ref{fig:WifiChannels} shows that even with a 20 MHz spread, channel 1 extends to approximately the centre of channel 3's frequency, while channel 6 spreads as far as channel 4, but the two do not overlap.

\subsection{MAC Addresses}
MAC stands for 'Media Access Control'  and is a unique 48 bit number assigned to network cards on their creation and is stored in read-only memory on the card to prevent modification. This number is used for interactions with other devices in a local network using Ethernet or WiFi. They are generally represented in six blocks of two hex digits (00:00:00:00:00:00). There are approximately 281 trillion valid MAC addresses.\\ \\ A useful feature of MAC addresses for this project is that to ensure each address assigned is unique, manufacturers are assigned blocks in the number space to use for their newly created devices. The first three blocks of hex in a MAC address are known as the Organisationally Unique Identifier (OUI), and are registered with manufacturers. These identifiers are updated regularly and can be found on the IEEE website. The benefit of this information was that it provided us with the manufacturer of the device that sent each detected probe request.


\section{Owner Identification}
The discretionary and exemplary challenges for this project related to attaching names, emails or some form of identifying information to MAC addresses scraped from probe requests. This means that the data set this project gathered needed to be compared with a secondary source containing personal information. Analysing these two data sets together lead to frequent co-occurrences between MAC addresses and users. After sufficient co-occurrences between a person and a MAC address it became apparent that this person was the owner of the co-occurring MAC address. 

\subsection{Challenges}
There were many challenges involved in this identification process. First and foremost is the fact that 	sensitive, identifying information was not accessible in the time frame for this project, primarily due to the fact that data protection laws would not permit unauthorised use of a data set by a third party (this project). Secondly, any potential data set needed to have a timestamp and location associated with them, the location could be inferred if the data was collected from a known location (such as a physical store).

Identifying users would be dependant on the amount of MAC addresses in the area of detection and whether target users tend to appear in the area in groups. If 300 MAC addresses are detected in an area while a single 	person appears in our secondary data set, it could take a significant amount of repeat appearances of 	this user to narrow these 300 MACs down to the one that belongs to them. Similarly, if this 	data set is from an activity people tend to do in groups (e.g. College Lectures) then a definite identification cannot be made if the same users and MAC addresses always appear simultaneously. 

\section{Secondary Datasets}
One of the core challenges involved in merging probe request data with a secondary dataset is identifying a suitable secondary dataset to make such an identification possible. Probe requests often contain sensitive information but they do not contain any definitively identifying information on their own. A secondary source of information is required to go from the data in probe requests alone to identifying a user by their MAC address. This section looks at some of the restrictions and challenges involved with finding and utilising a dataset that contains user information and correlates with probe requests in a manner that could be used to link this user information to detected MAC addresses. 

\subsection{Requirements}
First and foremost, the data set to be clustered with the probe request data needs to contain a method 	of communication with the user, with a contact mechanism this data is monetisable, without which 	the data is less significant as although it provides insight into a consumer base it does not provide an 	approach mechanism to reach these customers. Most likely, this data would contain an appropriate 	email address, phone number or social media account which would allow us to direct advertisements 	towards these identified users.  

In a perfect scenario, the instant a customer uses a company's service, the location and the time 	would be recorded and provided to the database, which would allow a correlation analysis to be made 	between the time and location these users interact with a company and the times and locations that 	various MAC addresses are detected.

\subsection{Collaboration Benefits}

A collaborator in the context of this project means any company with a suitable data set of consumers that would be willing to install probe detectors in their offices/stores to identify their customers by their MAC addresses. For many companies, knowing the instant a customer comes near one of their physical stores is 	valuable as it  is likely that the company can significantly increase its sales conversion rates if it 	targets marketing at customers only when they are within easy access of their store, but arguably the 	more significant benefit for a collaborator is that they would be entitled to a share in the sale and 	monetisation of the data they helped gather. Without the ability to approach nearby clients, this 	software is only applicable to be used as a form of people counter which can detect the rate of unique 	customers coming into the store as well as other interesting metrics such as the frequency of return 	visitors and the amount of return trips that users make, but with the identifying information, this data 	can be monetised and sold to a multitude of business's as a method of reaching customers with	customised marketing approaches that are currently within metres of a store; a highly valuable 	advertising market that is not currently offered by other advertising mediums. 

\subsection{Ethics/Legality}
Specific issues of legality are often difficult to answer when dealing with new technology as there isn't always a precedent to follow. The most appropriate document on the use of this identification software is by the Future of Privacy Forum (FPF)\cite{fpf}, a non-profit organisation working in the area of privacy in the emerging technologies sector. The FPF alongside leading companies in the area of mobile location analytics (MLA) developed a self-regulated "Mobile Location Analytics Code of Conduct" (Future of Privacy Forum, 2013)\cite{fpfmla} which would permit the use of this software under the following conditions:
\begin{itemize}
	\item That it was displayed to consumers in the area of detection that MLA was in use.
	\item That the software in use adhered to the centralised opt-out database for users not wishing to be profiled.
	\item That affirmative consent was received from customers to share this data with third parties.
	\item That third parties were contractually obligated to abide by the same code of conduct with regard to the data. 
\end{itemize}
Another source for guidance on the legality of this software comes from the United States Federal 	Trade Commission, who released what are now known as the Fair Information Practice Principles in 	an Electronic Marketplace (Federal Trade Commission, 2000)\cite{ftc} advising the US congress to legislate 	that all websites that gather personal identifying information about consumers should abide by four 	key principles that are:
\begin{itemize}
	\item Notice of the collection of data.
	\item The ability for a consumer to choose how their data is used.
	\item Reasonable access to the data held on a consumer.
	\item Safe storage of the data collected. 
\end{itemize}


\subsection{Approaches}
The most basic approach to identifying users would be to compare for co-occurrences of MAC addresses and users within a given time frame. For example, if given a set of all occurrences of a particular user in the secondary dataset, we generated a list of MAC addresses detected on the days the user was in the presence of our detector, we could analyse the lists for MAC addresses common to each day, if a particular MAC address appears the same day as a user appears, that increases the likelihood that the MAC belongs to the user, if they do not appear on the same day that decreases the likelihood of them belonging to each other. It is important to acknowledge that the absence of a MAC address on some occasions does not indicate that the pair are not linked as it is possible that the user did not have their device present at that time, or that the device was not transmitting probe requests for a variety of reasons. 

Improvements on this approach would be to cluster the data in such a way that the closer the 	detection time is to the secondary datasets recorded time, the higher the likelihood of a positive 	match, and the greater the time between the two events (including complete absence), the lower the 	likelihood of a positive match.

\section{Use Cases/Business Applications}
A large portion of the research for this project examined the potential applications of this data. What would be the implications of being able to generate a large dataset of MAC addresses linked with personal information (in particular, contact information)? The two main applications evident after research were in the areas of marketing and in national/cyber security. 

\subsection{Marketing}
Every modern company wants to personalise their approach to their customers to gain an edge on 	the competition, every modern company also wants to reach these customers in the perfect time and 	location, to maximise the chances of converting this interaction to a sale.

After initially developing a large dataset of MAC addresses linked to email addresses with a collaborating company, this dataset could be monetised under a B2B business model. A device as inexpensive as a Raspberry Pi could be placed near the front of a client's store detecting passing MAC addresses, if any that pass by are a customer of the client, the client could be made aware of the fact such that they could send them a personalised message/email, which would reach them while they are still in the immediate vicinity of the client's store; a level of business-consumer interaction that would be hard to beat.

The ability to deliver customised interactions to a customer when they are within 	touching distance of a store is highly valuable to the forward thinking advertiser. Information of this detail is currently possible through the use of apps with location tracking permissions authorised. However, according to research conducted by Urban Airship, opt in for location tracking and push notifications on apps are approximately 62\% and 51\% respectively\cite{urbanairship}, which also doesn't account for the difficulty in encouraging App downloads that most retailers face. MAC identification would not require an App download, nor location permissions to track the user's phone. A user would simply have to opt-in (with an explanation of the service) as they create an account with the company. 	A benefit to licensing this software is the fact that offering this service to new clients could include negotiating the use of their client data to identify even more MAC addresses as it would result in a shared benefit for 	both this software and the client looking to target even more of their customers, as not all of the clients customers would be identified in our dataset previously.

\subsection{Government Security}
Law enforcement departments the world over invest billions of dollars annually into security and 	surveillance systems. Everything from facial recognition systems to listening devices have been 	developed in an effort to cut crime around the world. The United States' surveillance budget is estimated at approximately \$75Bn annually\cite{cnnnsa} and the UK's new Investigatory Powers Act 2016\cite{ukinvestigate} has been estimated to cost the taxpayer over £11Bn to implement.\cite{investigatecost}

Many criminals in the modern age are very careful of where they go and what they do on the internet 	when they know they are wanted. What many don't know is that their phone or laptop, although  not being in use, is currently announcing its presence for anyone in the area to hear. If 	law enforcement were to get a hold of an email used by a criminal, or even better, a MAC address of 	one of their devices, every MAC sniffer running this software for companies looking to target their 	customers could also have the ability to alert the authorities of the exact time and location of the 	criminal if they were to detect the given address. This software would provide yet another medium 	for law enforcement to fight crime and search for wanted criminals. 

Consider the possibility that a probe detection system such as this was installed alongside a new public security camera system in a city, where sniffing devices were placed in public detecting MAC addresses. It is conceivable that detection of a MAC address could serve as evidence that the suspect's mobile device was present at the scene of a crime. Considering that the odds of two devices having the same MAC address by random chance is exceptionally small, it wouldn't be damning, but it would be difficult to dispute.

\chapter{MAC Address Identification}
This chapter documents the approach taken to merge simulated probe request data with a hypothetical dataset containing identifying information in an exploration into whether it is possible to link MAC addresses to personal information to strengthen the value proposition of collecting and analysing probe request data.

\section{Modelling the Identification Problem}
Early in this project it became apparent that it would not be feasible to get access to a dataset containing personal information on real people due to the sensitivity of such data. As a result this data was generated randomly in the form of a simulated experiment.

\subsection{The Simulated Experiment\label{experiment}}
There are many possible situations that would be suitable for identifying people by the MAC address of their devices, but the scenario had to be fixed to allow for a fairer comparison between identification approaches. The scenario that was modelled in this experiment was that of a college lecture, where each lecture had an attendance sheet and a sniffing device present that would detect all probe requests emitted during that lecture. 

This output of each simulated lecture resulted in two lists being generated per lecture, a list of names of attendees and a list of detected MAC addresses. The identification algorithm took these lists on a lecture by lecture basis and tried to find a correlation between users and MAC addresses to infer that a given MAC address belonged to a given user. The variables controlled in this simulation were as follows;

\begin{itemize}
	\item The attendance probability - The probability that a given person was going to attend the given lecture.
	
	\item The probability a given device emits a probe request - Different devices emit probe requests at different intervals, Julien Freudiger\cite{freudiger} analysed three popular devices and found that all devices emitted probe requests at least every 330 seconds, but the algorithm was tested under the assumption that not all devices emit probe requests.  
	
	\item The number of lectures - With more detection opportunities a more accurate assignment of MAC addresses is possible. The number of lectures simulated was controlled to analyse how the identification algorithm performed in sub optimal cases as well best case scenarios. 
\end{itemize}

\subsection{The Occupancy Grid Mapping Algorithm\label{occupancy}}
The basis for the identification algorithm was developed from occupancy grid mapping algorithms. Occupancy grid mapping is an algorithm commonly used by mapping devices and robotic vision first developed in 1985 (Moravec \& Elfes, 1985).\cite{moravec} The aim of the algorithm is to identify objects in a space using an imperfect detection method that is subject to noise. A zero matrix is generated for the area to be scanned, each cell contains a probability that the cell is occupied. Continual scanning is performed across all cells and the results for each successive scan either increases or decreases the probability of a given cell being occupied. 

This translates to the MAC address identification problem as such. The algorithm generates a 2D array the size of the number of names and number of detected MAC addresses. As the algorithm is passed lists of attendees and detected MAC addresses, it increases and decreases the weights of each cell accordingly. A visualisation of this process is below. 
\begin{figure}[H]
	\centering
	\includegraphics[width=10cm]{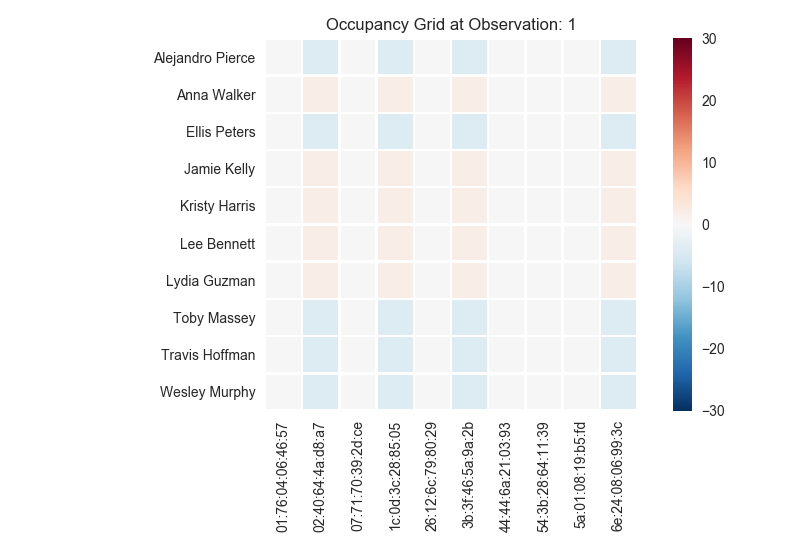}
	\label{fig:occupancy1}
	\caption{Occupancy grid after one lecture}
\end{figure}
\begin{figure}[H]
	\centering
	\includegraphics[width=10cm]{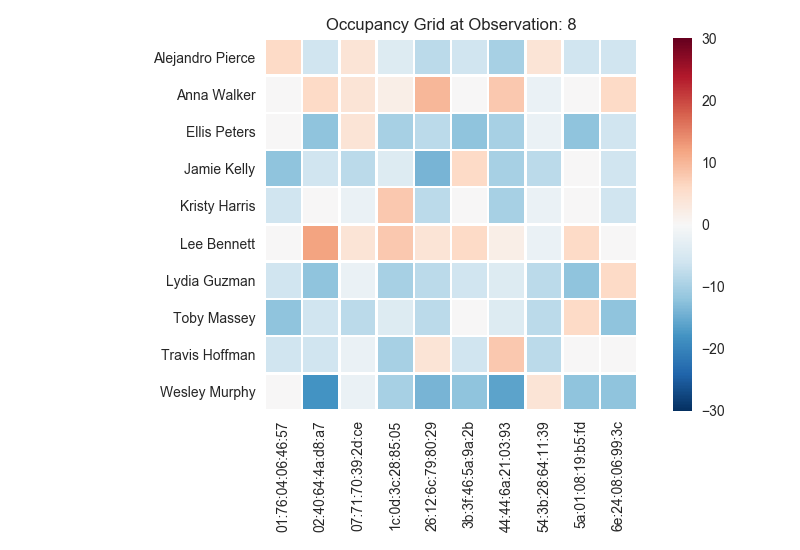}
	\label{fig:occupancy8}
	\caption{Occupancy grid after eight lectures}
\end{figure}
\begin{figure}[H]
	\centering
	\includegraphics[width=10cm]{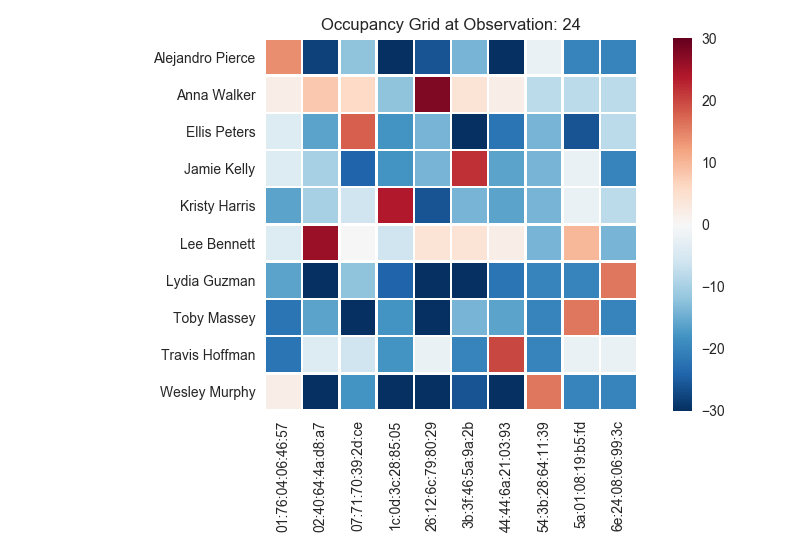}
	\label{fig:occupancy24}
	\caption{Occupancy grid after twenty four lectures}
\end{figure}

In the outlined experiment there are four possible scenarios for a given MAC address/Name pair; The person is present, the MAC isn't, the MAC is present, the person isn't, neither are present or both are present. The following table indicates whether the given scenario increases or decreases the probability that this MAC address belongs to this person. 

\begin{centering}
	
	\begin{tabular}{ |p{3cm}|p{3cm}|p{3cm}| }
		\hline
		& Person Present 	& Person Absent	\\
		\hline
		MAC Present		&$\uparrow$			&$\Downarrow$		\\
		\hline
		MAC Absent		&$\downarrow$		&- 				\\	
		\hline
	\end{tabular}
	
\end{centering}

\section{Algorithm Evaluation}
From first implementation on, the identification algorithm was successful in identifying people provided the amount of data points was relatively high. It quickly became a question of how little data can the algorithm perform effectively on, and which influencing factors affect the identification process the most. 

\subsection{Correctness}
For the purpose of tracking how the identification algorithm performed as it learned more and more about the target users, the algorithm attempted to assign a MAC address to every user after each new 'lecture' (titled Observations in the graphs). In a real scenario, a confidence threshold would be implemented to only identify users which passed a certain threshold. Similarly, the evaluation of different algorithm implementations only considered total correct assignments, and did not penalise algorithms that made many incorrect guesses either. 

\subsection{Statistical Analysis of the Simulation Parameters}
After choosing an optimal identification algorithm, regression analysis was performed on the independent variables (as discussed in Section \ref{experiment}) to identify which variables had the most influence on the accuracy of the algorithm. The following table is a summary of the regression statistics for almost 500 simulations with varied input parameters. 

\begin{centering}
	
	\begin{tabular}{ |p{3cm}|p{3cm}| }
		\hline
		\multicolumn{2}{|c|}{Regression Summary}	\\
		\hline
		Multiple R			&0.88036614			\\
		\hline
		R Square			&0.77504454			\\	
		\hline
		Adjusted R Square	&0.77359009			\\	
		\hline
		Standard Error		&0.03119968			\\	
		\hline
		Simulations			&468				\\	
		\hline
	\end{tabular}
	
\end{centering}

\begin{figure}[H]
	\centering
	\includegraphics[width=14cm]{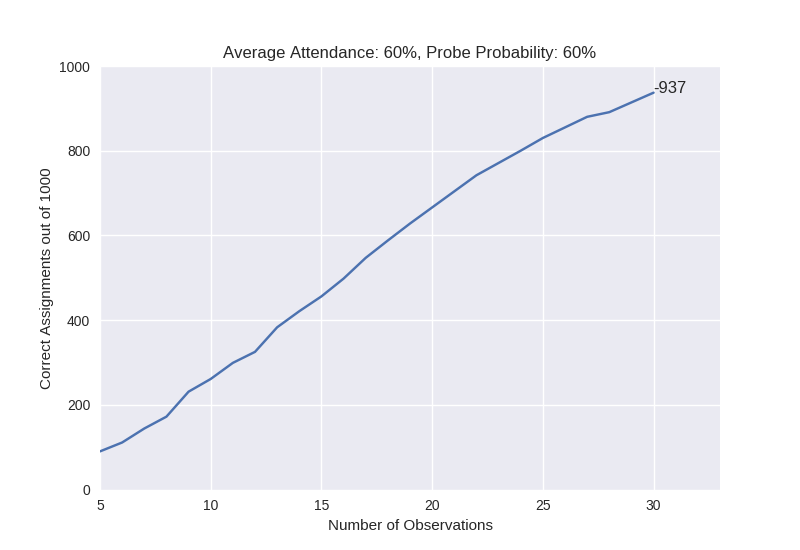}
	\caption{Correctly identified users over the span of 30 observations with a low probe request probability}
	\label{probe60}
\end{figure}

\begin{figure}[H]
	\centering
	\includegraphics[width=14cm]{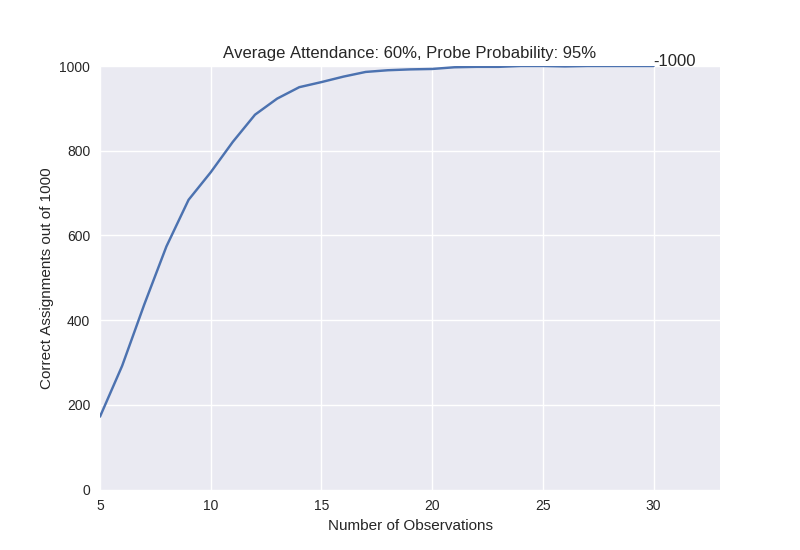}
	\caption{Correctly identified users over the span of 30 observations with a high probe request probability}
	\label{probe95}
\end{figure}

An adjusted R square value of 0.77 indicates that there is a significant positive correlation between the three independent variables (Observation Number, Probe Probability, Average Attendance) and the dependent variable, accuracy. All three variables had a P-value under 0.05 which indicates that they are statistically significant at a 95\% confidence interval. The probe probability variable had a correlation coefficient of approximately 0.58, controlling for the other variables. This indicates that the accuracy of our identification algorithm was strongly linked to the detection of probe requests. This is evident when graphing the accuracy of the algorithm over the course of a set of observations. Figure \ref{probe60} shows the accuracy of the identification algorithm when 60\% of devices emitted probe requests. While figure \ref{probe95} shows the significant improvement in accuracy of the algorithm when 95\% of devices emit probe requests. 

After this evaluation it is apparent that identification of people based on co-occurrences between their attendance at a known location and detection of their MAC addresses at the same location is definitely possible and the challenge is to minimise the amount of detections required to make an accurate identification. If the granularity of the detection was reduced such that rather than having a list of all attendees and a list of all detected MAC addresses the dataset included exact arrival times for people and first detection times for MAC addresses then accuracy could be improved even further but this was beyond the scope of the experiment. 

\chapter{Probe Request Capture \& Analysis}
After analysing the effectiveness of the identification algorithm in matching MAC addresses to personal data, the second aspect of this project revolves around capturing these MAC addresses rather than simulating them. This chapter looks at detecting and parsing probe requests, extracting MAC addresses along with any other personal information from them and storing this information in a database. 

The following is a brief overview of the probe request capture process followed by an in depth look at each stage of the system;

\begin{itemize}
	\item A dedicated laptop was deployed in a public location to periodically record probe requests of nearby devices.
	
	\item This device executed a shell script every 15 minutes. This script controlled the detection of probe requests, the parsing of this data and the upload of this data to the database, which was stored on a web server to allow for multiple sniffing devices to operate simultaneously.
	
	\item When the server received an upload of new MAC addresses it either created database entries for new MAC addresses or updated existing entries to reflect the new detections of existing MAC addresses. 
	
	\item A web dashboard was developed to display a high level overview of this data including dynamic pages that can be generated for each MAC address.
	
	\item Web pages for individual MAC addresses plot all GPS locations where we've detected the MAC address and in cases where SSIDs have a known GPS location, those SSIDs are also plotted. Both the general dashboard and the individual MAC profile dashboards are below.

\end{itemize}

\begin{figure}[H]
	\centering
	\includegraphics[width=14cm]{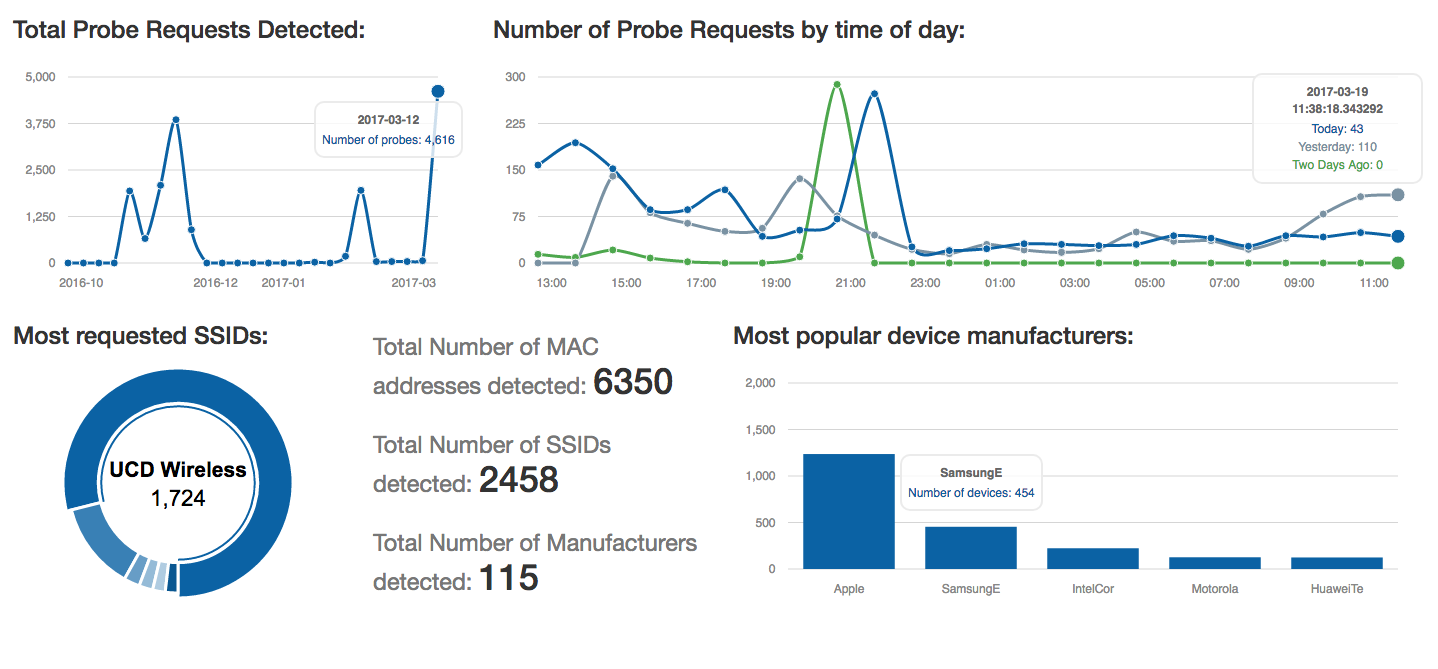} \\
	\caption{The global dashboard logging all probe requests recorded.} 
	\label{dashboard}
\end{figure}

\begin{figure}[H]
	\centering
	\includegraphics[width=14cm]{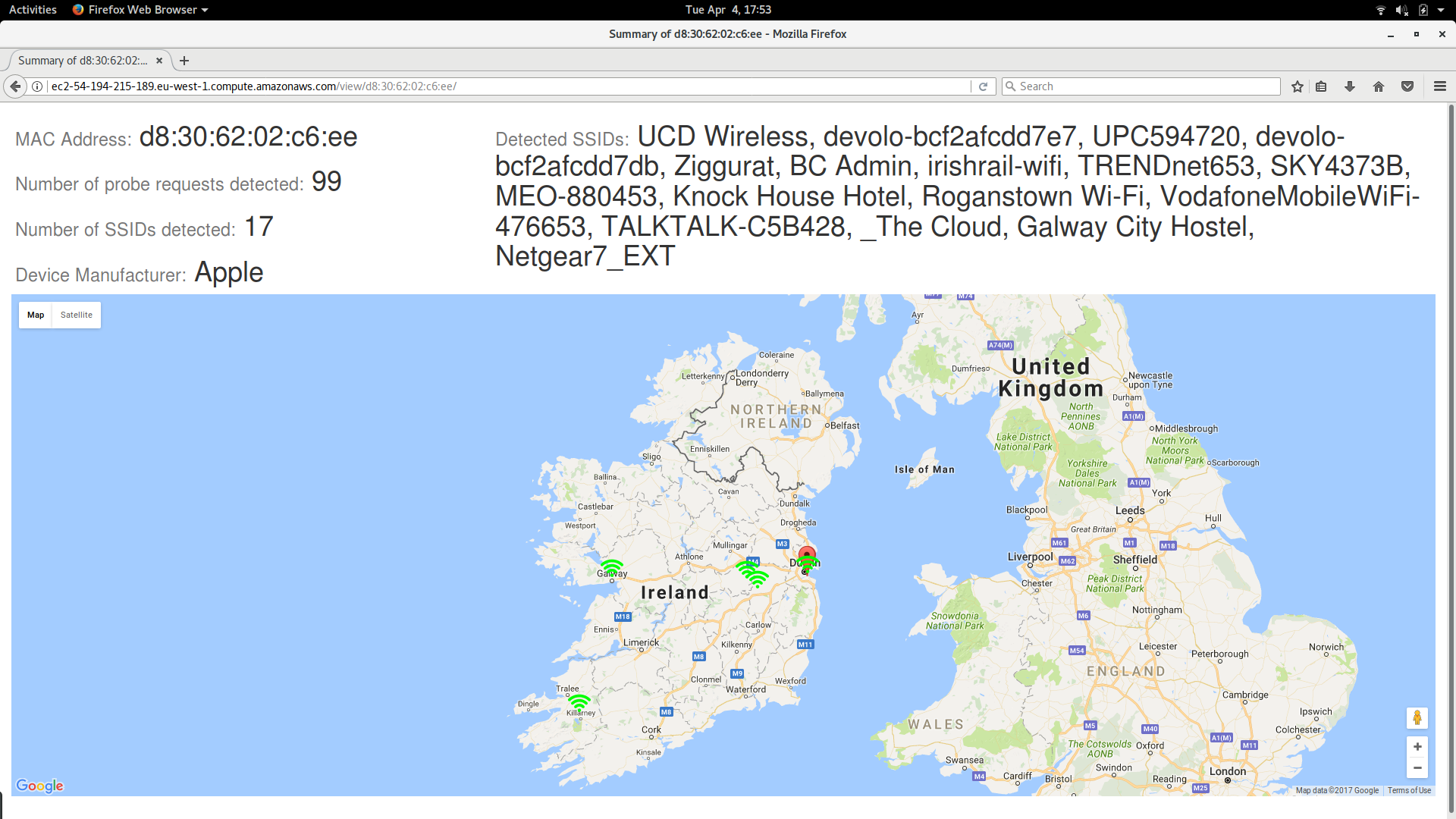} \\
	\caption{A custom dashboard generated for a detected MAC address.} 
	\label{macdashboard}
\end{figure}

\section{Sniffing Devices}
A self-imposed requirement for this probe request capture software was that any probe request capture software should be scalable. As such the system was developed with the intention of being distributed.  Multiple sniffing devices are capable of being in operation simultaneously detecting probe requests in different locations and all contributing to a single global database of MAC addresses. 

\subsection{Sniffing Device Overview}

The command line tool WireShark uses for sniffing on Unix systems is called dumpcap\cite{dumpcap}. Dumpcap was used as the core tool for detecting probe request packets in this project. A dedicated laptop had a cron job execute a sniffing script every 15 minutes. This script put the device into monitor mode, detected probe request packets for 45 seconds before parsing these packets to store each MAC address and any SSIDs they may have searched for before uploading them to the probe request database. 

\subsection{Data Size}
Data size was of significant interest when developing the data gathering solution. A significant amount of testing was done to estimate the amount of probe request packets picked up by a probe detector in a given period of time. 

After some exploratory testing, approximately 500 probe requests yielded a JSON file of 	parsed data of approximately 50Kb in size which, after some optimisations, was reduced to under a one second upload to 	the centralised detection database. A crowded area can yield up to 10 probe requests per second. 

Further into the development of this project it became apparent that probe requests tend to be sent in bursts across multiple channels successively. This meant that there was a significant amount of redundant data being uploaded to the server of multiple identical probe requests spaced milliseconds apart. In an effort to reduce this redundancy, improvements were made to the parsing of the detected probe requests such that only one probe request per mac address per SSID was saved after each scan. Meaning a MAC address that only used the broadcast address was saved once, and MAC addresses that searched for specific SSIDs had one request per SSID saved. This reduced the amount of probe requests being stored by roughly a factor of eight.\\

By the end of this project, 17,511 probe requests have been recorded coming from 6,926 unique MAC addresses manufactured by 120 different manufacturers.

\subsection{SSIDS}
Over the course of this project if any probes were directed at a specific SSID rather than the broadcast address, the SSIDS were stored to a database and linked by a many to many relationship to any MAC address that has searched for it. At the time of writing this report over 500 unique SSIDs have been detected.\\

\begin{figure}[H]
	\centering
	\includegraphics[width=9cm]{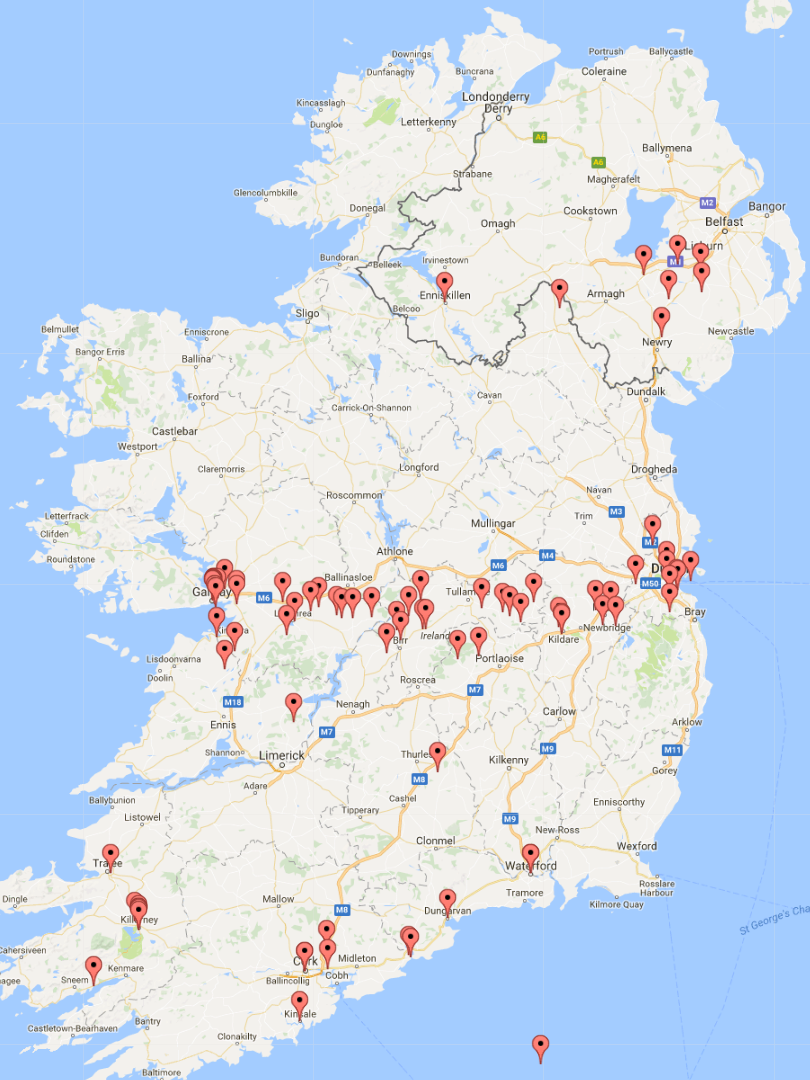}
	\caption{SSID Locations}
	\label{SSID:Locations}
\end{figure}

With the use of Python scripts by project supervisor, Dr. MacNamee and data scraped from wigle.net\cite{wigle}, GPS locations were approximated for 87 of  these SSIDS. Fig \ref{SSID:Locations} plots the locations of each generated SSID using software developed by Darrin Ward. \cite{darrinward}

\section{Probe Request Database}

The centralised server built to store probe requests from all sniffing devices was developed using the Django Web Framework\cite{django}. The database was implemented in SQLite. The server was hosted on the AWS cloud. A simple REST API was developed to accept the JSON representation of the sniffed data, the server parses this JSON and stores each probe request as a sighting, adds each MAC address to the MAC table, and any SSIDs to the SSID table. The owner table is never automatically updated as the identification process is only performed on generated data rather than real probe requests. Below is a schema diagram for the Database. 

\begin{figure}[H]
	\includegraphics[width=\linewidth]{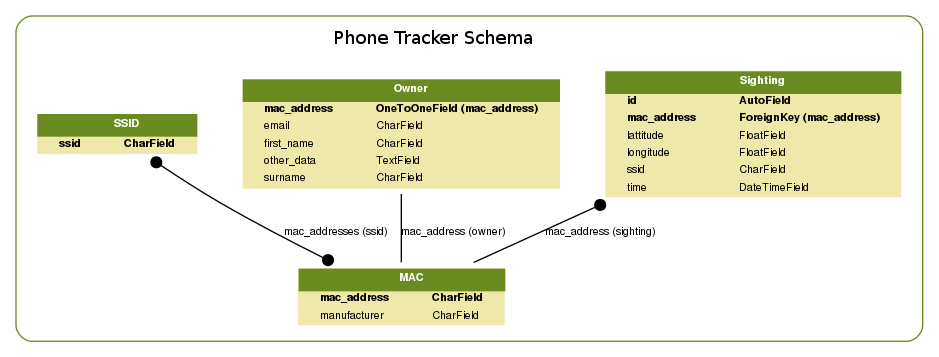}
	\caption{The Database Schema}
	\label{fig:schema}
\end{figure}

\chapter{Detailed Design \& Implementation}

\section{Sniffing Device}
The most popular network sniffing software on the market is WireShark\cite{wireshark}. Early initial research in this 	project was conducted through the application as it excels in providing a detailed filtering and 	breakdown of packets detected; it is also able to read packets detected by other software in the 	format .pcap. 

The following is a more detailed overview of the sniffing device's implementation than the outline in previous chapters.

\begin{itemize}
	\item A cron job was used to execute a custom shell script every 15 minutes. 
	
	\item First, the Dumpcap tool was used to put the device into monitor mode (section \ref{monitor}) to record all probe request packets for 45 seconds. These packets were stored in .pcap files. Dumpcap was used for this task as it can be supplied with filter arguments to filter out all packets except probe requests. 
	
	\item Next a Python script reads this file using the ScaPy package\cite{scapy} and creates a dictionary of MAC addresses and the associated probe requests they sent. Devices send probe requests in bursts across many WiFi channels so to reduce the amount of redundant data being stored, for each MAC address only one probe request was stored after each scan unless the probes were directed at a specific SSIDs. This reduced the amount of data being stored by approximately a factor of eight.
	
	\item This data was converted to JSON and uploaded to the probe request server via a POST request. The uploaded content looked like this:
		\begin{verbatim}
		{
		"80:7a:bf:3b:bd:d9":{
		"detections":[
		["2016-12-04 19:23:54","UCD Wireless",[53.3461, -6.3032]],
		["2016-12-04 19:24:32","Broadcast",[53.3461, -6.3032]]]],
		"times_seen":2,
		"manufacturer":"Apple"}
		}
		
		\end{verbatim}

	\item This data is sent to the probe database using a Python package called Requests\cite{python:requests}.

\end{itemize}

\subsection{Timing\label{timing}}
One minor issue with the implementation used in this project is that the packets detected initially by the dumpcap tool were not immediately timestamped. The timestamps were added after collection while they were parsed for upload. This means that some probe requests can have their time stamp off by up to 45 seconds (the scan length for the sniffing devices). This was not viewed as a significant issue in the context of this project. If the opportunity arose to redesign the sniffing devices, dumpcap would not be chosen for this reason, instead, ScaPy, the python package used to parse the dumpcap output, could have been used to detect the packets and immediately timestamp them all in one. 

\section{Android}
One major question not addressed in this report so far is whether this technology could be implemented for a mobile device rather than for a Unix system. In general, most Unix devices are laptops or PCs and do not contain GPS receivers which are standard in modern smartphones, so they need to have the GPS coordinates of their location manually input. Unfortunately, after a significant amount of research into the topic it has become apparent that only a tiny subset of Android devices would be capable of detecting probe requests. These devices would need to be have root permissions to install custom kernel patches to make probe request detection possible. The main problem with developing a solution for Android is the fact that the wireless chipsets on almost all smartphones do not support monitor mode. The two potential implementations of Android sniffing devices are listed below but development of these solutions was not pursued in this project due to the relatively few cases where such a situation would work. 

\begin{itemize}
	\item Some Android devices support a USB standard called USB OTG\cite{symlis} which allows the phone to control USB peripherals like a normal PC would. If these devices were rooted, then custom software could be written to interact with a USB network interface card capable of monitor mode. 
	
	\item Alternatively, a small number of Android devices (Samsung Galaxy S1, S2, Nexus 7 and Huawei Honor)\cite{aircrackandroid} have been developed using the Broadcom 4329 and Broadcom 4330 chipsets which natively permit monitor mode. Any of these devices could theoretically be rooted and custom software could be written to convert these devices into probe request detectors without an external network card. 
\end{itemize}

\section{Probe Database}
Hosting a centralised database was important to demonstrate the scalable nature of a probe request sniffing solution. The first implementation consisted of a Django web server running locally on the same device as the probe detector. After developing this system locally the server needed to be moved to a web server accessible over the internet. The first approach was to use Heroku, for it's free hosting and easy deployment by tracking the master branch of the server's git repository and automatically pulling and deploying any new pushes to the master branch. Unfortunately, Heroku uses an ephemeral file system rather than a disk backed system so new database entries were wiped every hour. To solve this either the database would need to be moved to PostgreSQL or a different cloud provider would need to be used. The decision to move providers rather than changing the database was made because at this point in time, the plan to develop the probe detection system for Android was still in place and Android uses SQLite as a database so it seemed appropriate that the global database remained in SQLite. 

Similar issues were encountered while hosting the server on Amazon's Elastic Beanstalk platform so finally the third approach of spinning up a dedicated VM on AWS's EC2 cloud was used. This proved to be an efficient solution as the Linux based VM needed similar configuration to the local Linux based server and all of the files could easily be moved from the local server to the production server via commits and pulls to the remote git repository for the server.

\section{Web Dashboards}
Two important tasks in the project brief were to;

\begin{itemize}
	
	\item Visualise statistics on probe requests collected in a real-time dashboard

	\item Profile people by using data associated with probe requests, or other connected data sources.
	
\end{itemize}

Online web dashboards were developed as solutions to both of these tasks, as presented in Fig. \ref{dashboard} and Fig. \ref{macdashboard}. These dashboards combined the use of the Bootstrap Framework\cite{bootstrap} and MorrisJS\cite{morris}, a JavaScript charting library. Bootstrap was used to develop responsive web layouts for each graph and chart on both desktop and mobile browsers. MorrisJS was used to visualise the information from the probe request database. Django's template system made it possible to substitute Python data structures with placeholders in the dashboard's html in real time for MorrisJS to use for it's visualisation. 

\section{Simulated Identification}
The simulated identification experiments in this project were written in Python using an Object Oriented approach. The simulation process was as follows: 
\begin{itemize}
	\item The two key objects were the Experiment and the Observer, the experiment object took a list of random names, generated random MAC addresses for each and stored them in a file as the true solution. 
	
	\item Next, for each lecture, it used the attendance and probe probability values to randomly create the attendance list and MAC address list which it passed to the Observer object.
	
	\item The observer object attempted to predict which MAC address belonged to each name using the occupancy grid mapping algorithm described in Section \ref{occupancy}. 
	
	\item The observer object would pass a proposed assignment of names to MAC addresses back to the Experiment object, the experiment object would then rate the accuracy of the proposed assignment.
	
	\item Matplotlib was used to visualise the accuracy of these assignments and these visualisations can be seen in Fig. \ref{probe60} and Fig. \ref{probe95}.
\end{itemize}

Some modifications to the algorithm were made to improve the algorithm performance. These modifications include:
\begin{itemize}
	\item In the interest of realism, the entire list of names is supplied to the Observer object on creation, but it generates a list of MAC addresses dynamically as they appear in observation lists as these would not be known in advance of the experiment in a real world scenario. These MAC addresses are only linked to names that co-occur with them rather than being applied to all people. 
	
	\item After every new observation, the algorithm prunes the least likely MAC addresses from the potential candidates to prevent the grid becoming excessively large in crowded public areas with many detected MAC addresses. 
\end{itemize}

\subsection{Algorithm Evaluation}
Each time a person attends a lecture, their candidate MAC addresses need to be updated with new probabilities. Because the algorithm caps the maximum amount of MAC addresses that can be linked to each person, the algorithm performs a constant amount of work per person. This work is performed for each person that attends a lecture. Every experiment consists of multiple lectures. As a result, the overall performance of this algorithm is $O(n*m), n := \vert Lectures\vert , m := \vert People\vert$

\begin{figure}[H]
	\centering
	\includegraphics[width=12cm]{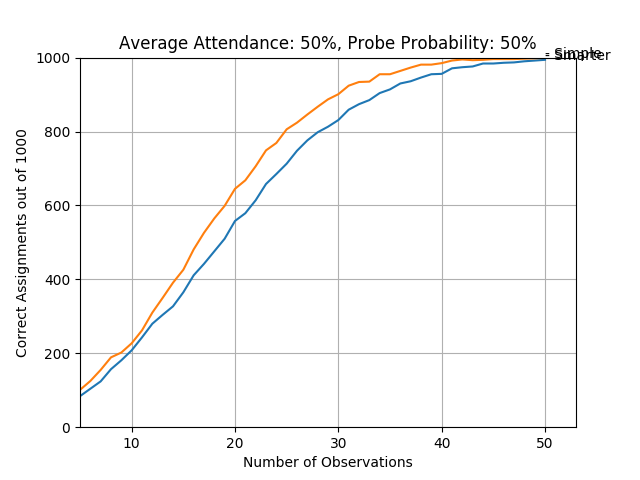}
	\caption{Orange indicates duplicate assignments, Blue indicates unique assignments}
	\label{uniqueassignments}
\end{figure}

An interesting aspect to the performance of this algorithm displayed in Fig. \ref{uniqueassignments} is that insisting that different MAC addresses are assigned to different people noticeably reduces the accuracy of the algorithm. Allowing the same MAC address be assigned to multiple people ensures that at least one of those assignments is correct whereas insisting that each assignment is unique reduces this accuracy. This situation would not be as significant an issue in an implementation that only assigns MAC addresses over a certain confidence threshold. 

\chapter{Conclusions \& Future Work }

\section{Conclusions}
This project attempted to prove two things; that people could be identified by their MAC addresses given a secondary dataset containing personal information, and that these MAC addresses can be gathered in a manner that would make this identification possible. At the conclusion of this project I feel like both of these tasks are absolutely possible and I think this project sufficiently demonstrates this. The probe request capture software is live and recording real MAC addresses along with their locations and the simulated identification algorithm identifies people with high levels of accuracy given a sufficient amount of data. Of course it would have been ideal to perform this identification with a real data set of people but unfortunately getting access to a data set of this information was not achieved.

Upon reflection on the approaches and decisions made in this project, I am happy with the results of most. The two main aspects that I would change were I to begin again would be to use ScaPy alone to detect and parse the probe request packets to resolve the timing error discussed in section \ref{timing} and I would develop a simulation with a smaller level of granularity than the current one, where each person had an arrival time recorded and each probe request had a first detection time. Such that a probability density function could be used to weight MAC addresses according to a normal distribution curve centred around the target users arrival time. 

All things considered I could not be more happy with the promising results of the project and I consider the project to be a success as a whole. 

\section{Future Work}
If this project were to be developed further, the proposed changes mentioned above with regards to the sniffing solution and the more complex simulation set up would first be implemented and a redesigned identification algorithm would be developed. After this, the grading of algorithm accuracy would be made more punitive to reduce the number of incorrect assignments made such that a good identification algorithm must implement a confidence threshold on it's candidate assignments. 

This would result in algorithms being compared on both their accuracy and their coverage, rather than simply accuracy alone. Other identification approaches could be explored including clustering algorithms drawn from data mining research and correlation analysis based on a more statistical approach to the identification process. 

Of course if this project were to be developed further, every effort would be made to obtain or create a data set of personal information to link to detected MAC addresses, to test the identification algorithms on real data rather than simulated data. 



\label{endpage}


\newpage
\begin{thebibliography}{99}
	\bibitem{freudiger}Freudiger, J. (2015) How Talkative is your Mobile Device?
	An Experimental Study of Wi-Fi Probe Requests. Available at: https://frdgr.ch/wp-content/uploads/2015/06/Freudiger15.pdf
	
	\bibitem{wigle}Wireless Geographic Logging Engine: https://wigle.net/
	
	\bibitem{802.11}Institute of Electrical and Electronics Engineers (2012) Telecommunications and information exchange between systems.
	Local and metropolitan area networks - Specific requirements Part 11: Wireless LAN Medium Access Control
	(MAC) and Physical Layer (PHYS) Specifications. http://standards.ieee.org/about/get/802/802.11.html
	
	\bibitem{gast}Gast, M.S. (2005) 802.11 wireless networks: The definitive guide. 2nd edn. Boston, MA, United States: O?Reilly Media, Inc, USA.
	
	\bibitem{metageek} MetaGeek: http://www.metageek.com/training/resources/why-channels-1-6-11.html
	
	\bibitem{fpf}Future of Privacy Forum: https://fpf.org/
	
	\bibitem{fpfmla}Future of Privacy Forum (2013) Mobile Location Analytics Code of Conduct. Available at: https://fpf.org/wp-content/uploads/10.22.13-FINAL-MLA-Code.pdf (Accessed: 5 November 2016).
	
	\bibitem{ftc}Federal Trade Commission (2000) Privacy Online: Fair Information Practices In The Electronic Marketplace; A Report To Congress. Available at: https://www.ftc.gov/sites/default/files/documents/reports/privacy-online-fair-information-practices-electronic-marketplace-federal-trade-commission-report/privacy2000.pdf (Accessed: 5 November 2016).
	
	\bibitem{urbanairship} Urban Airship, Location and Push Notification Opt-In blog post: https://www.urbanairship.com/blog/location-opt-in-rates-show-bright-future-for-ibeacon
	
	\bibitem{cnnnsa}CNN Money (2013) What the NSA costs taxpayers: http://money.cnn.com/2013/06/07/news/economy/nsa-surveillance-cost/
	
	\bibitem{ukinvestigate} Investigatory Powers Act 2016:\\ http://www.legislation.gov.uk/ukpga/2016/25/contents/enacted
	
	\bibitem{investigatecost} Full Fact(2012) The "Snooper's Charter": What's the price of listening in? https://fullfact.org/crime/snoopers-charter-whats-price-listening/
	
	\bibitem{moravec}Moravec, H. and Elfes, A. (1985). High resolution maps from wide angle sonar. Proceedings. 1985 IEEE International Conference on Robotics and Automation.
	
	\bibitem{wireshark}Wireshark: https://www.wireshark.org/
	
	\bibitem{dumpcap}Dumpcap: https://www.wireshark.org/docs/man-pages/dumpcap.html
	
	\bibitem{scapy}Scapy, by SecDev.org: http://www.secdev.org/projects/scapy/
	
	\bibitem{django}The Django Project: https://www.djangoproject.com/
	
	\bibitem{bootstrap} Bootstrap, the HTML, CSS and JS Framework. http://getbootstrap.com/
	
	\bibitem{morris}MorrisJS, http://morrisjs.github.io/morris.js/
	
	\bibitem{Android:2015} Android 6.0 changes (no date) Available at:\\ https://developer.android.com/about/versions/marshmallow/android-6.0-changes.html\#behavior-hardware-id (Accessed: 16 November 2016).
	
	\bibitem{python:requests}Requests - HTTP for Humans: http://docs.python-requests.org/en/master/
	
	\bibitem{symlis} List of Android devices supporting USB OTG - https://www.symlis.com/blog/2015/1/12/list-of-otg-supported-devices.
	
	\bibitem{aircrackandroid} Aircrack Android. An android port of Aircrack-ng. https://github.com/kriswebdev/android\_aircrack/blob/master/README.md
	
	\bibitem{darrinward}Darrin J. Ward, Plot Lat/Long Points on a Map: http://www.darrinward.com/lat-long/
	
\end{thebibliography}
\end{document}